\documentclass[preprintnumbers, prd, twocolumn, floatfix,
preprintnumbers,showpacs, nofootinbib]{revtex4}
\usepackage{graphicx,epsfig}
\usepackage{amsmath}
\usepackage{amssymb}
\usepackage{color}
\usepackage{amsfonts}
\begin{document}
\title{ Light propagation and optical scalars in torsion theories of gravity }
\author{S. Akhshabi}
\email{s.akhshabi@gu.ac.ir}
\affiliation{Department of Physics,
Golestan University, P. O. Box 49138-15759\\ Gorgan, IRAN}
\begin{abstract} We investigate the propagation of light rays and
evolution of optical scalars in gauge theories of gravity where
torsion is present. Recently the modified Raychaudhuri equation in
the presence of torsion has been derived. We use this result to
derive the basic equations of geometric optics for several different
interesting solutions of the Poincar{\'e} gauge theory of gravity.
The results show that the focusing effects for neighboring light
rays will be different than general relativity. This in turn has
practical consequences in the study of gravitational lensing effects
and also determining the angular diameter distance for cosmological
objects.
\end{abstract}
\maketitle
\section{Introduction}
In his pioneering paper of 1955, Raychaudhuri derived his now famous
equation for the volume expansion rate for congruences of timelike
geodesics \cite{Ray}. The main motivation of that paper was to
include vorticity and shear in the description of the cosmic fluid
(\emph{i.e.} relaxing the assumptions of homogeneity and isotropy)
and to study whether it can lead to non-singular cosmological
solutions. Later, the equation and its consequences were extensively
used by Penrose and Hawking in their development of the singularity
theorem \cite{Pen,Hawk}.

An extension of the Raychaudhuri equation to null geodesic
congruences were first derived by Sachs in 1961 \cite{Sachs}. The
reader can find a vigourous mathematical derivation of this equation
in general relativity in section 4.2 of reference \cite{Hawk2} and
also in \cite{Wald} . This equation can be used in studying light
propagation in various cosmological backgrounds and also for the
study of gravitational lensing \cite{Sasaki,SEF}.

All of the above studies have been done in a Riemannian background
spacetime where the torsion tensor vanishes, however it has been
shown that in a spacetime with non-vanishing torsion,  the
kinematical quantities describing the flow of the cosmic fluid,
\emph{i.e.} shear, rotation, and volume expansion rate and the
equations governing their evolution should be modified
\cite{Puetz,Yasskin,Nom,Cap}. This naturally leads to a
generalization of the Raychaudhuri equation in presence of torsion
leading to a more general understanding of the phenomenon of
geodesic focusing.

On the other hand, torsion appears naturally in many gauge theory
descriptions of gravity. The main idea of these theories is that
global symmetries are not compatible with field theoretical
description of nature and as a result these symmetries should be
replaced with local ones. If we apply this localization scheme to
the symmetry group of special relativity, \emph{i.e.} global
Poincar{\'e} transformations which include both Lorentz rotation and
translation, and demand that the Lagrangian of the theory remain
invariant under new local transformations, we arrive at Poincar{\'e}
gauge theory of gravity (PGT) in which both curvature and torsion
are present \cite{BLAG,Hayashi}. The dynamical variables in this
theory are tetrad and spin connection fields and the field strengths
associated with these fields are curvature and torsion tensors.
Curvature and torsion are coupled to energy-momentum and
spin-density tensors respectively. Poincar{\'e} gauge theory of
gravity contains some important special cases: General relativity
(vanishing torsion), teleparallel theory (vanishing curvature) and
also Einstein-Cartan theory where the Lagrangian is set to be equal
to the Einstein-Hilbert Lagrangian of general relativity.
Einstein-Cartan theory offers the simplest generalization of general
relativity and has been studied extensively in literature
\cite{Kerlick}. In this case the torsion is completely determined by
the spin density tensor and can not propagate \cite{Rauch}. However,
propagating torsion modes can be present in Poincar{\'e} gauge
theory of gravity with general quadratic Lagrangian \cite{Blag2}.

The organization of the paper is as follows: in section II we
present the recently derived Raychaudhuri equation in the presence
of torsion for null geodesics, section III is a brief review of the
Poincar{\'e} gauge theory of gravity including the definition of the
dynamical variables, most general form of the Lagrangian and also
the field equations. The acceptable forms of the torsion and spin
density tensors for a homogeneous and isotropic cosmological
background are also presented in this section. In section IV we use
the Raychaudhuri equation to study the light propagation and
evolution of optical scalars in Poincar{\'e} gauge theory of gravity
and derive the exact relation for the optical scalars for two
different interesting cosmological solutions of the Poincar{\'e}
gauge theory of gravity. Section V is devoted to conclusion and a
brief discussion of the main results.

\section{Raychaudhuri equation for null geodesics and its extension to spacetimes with torsion}
In this section we give a brief review of the Raychaudhuri equation
for null geodesics in general relativity and also its extensions to
spacetimes with torsion \cite{Poisson}. We begin by considering a
congruence of null geodesics which are affinely parameterized and
with a tangent vector field denoted by $k^{\alpha}$. We also
introduce the deviation vector $\xi^{\alpha}$ representing the
separation between corresponding points (same values of the affine
parameter) on neighboring curves. With this definitions we have the
following equations

\begin{equation}
k^{\alpha}k_{\alpha}=0 \quad k_{~~;\beta}^{\alpha}k^{\beta}=0 \quad
k_{~;\beta}^{\alpha}\xi^{\beta}=\xi_{~~;\beta}^{\alpha}k^{\beta}
\quad k^{\alpha}\xi_{\alpha}=0
\end{equation}
Where the semi-colon denotes the covariant derivative. In order to
isolate the purely transverse part of the deviation vector, we need
to introduce another auxiliary null vector field
 $N^{\alpha}$ with the following properties
 \begin{equation}
N^{\alpha} N_{\alpha}=0 \quad N^{\alpha} k_{\alpha}=-1 \quad
N_{~~;\beta}^{\alpha} k^{\beta}=0
\end{equation}
Using this vector, the transverse metric is then given by
\begin{equation}
h_{\alpha\beta}=
g_{\alpha\beta}+k_{\alpha}N_{\beta}+N_{\alpha}k_{\beta}
\end{equation}
In the next step we introduce the tensor field
\begin{equation}
B_{\alpha\beta}=k_{\alpha;\beta}
\end{equation}
The purely transverse part of $B_{\alpha\beta}$ is
\begin{equation}
\tilde{B}_{\alpha\beta}=h_{~\alpha}^{\mu}h_{~\beta}^{\nu}B_{\mu\nu}
\end{equation}
where the ' $\tilde{}$ ' denotes the transverse part of the tensor.
In this setup, the vector
$\tilde{B}^{\alpha}_{~~\beta}\tilde{\xi}^{\beta}$ can be regarded as
the transverse relative velocity between two neighboring geodesics.
$\tilde{B}_{\alpha\beta}$ can be decomposed into its irreducible
parts as follows
\begin{equation}
\tilde{B}_{\alpha\beta}=\frac{1}{2}\theta
h_{\alpha\beta}+\sigma_{\alpha\beta}+\omega_{\alpha\beta}
\end{equation}
where the kinematic quantities
\begin{equation}
\theta=\tilde{B}_{\alpha}^{\alpha} \quad ,
\sigma_{\alpha\beta}=\tilde{B}_{(\alpha\beta)}-\frac{1}{2}\theta
h_{\alpha\beta} \quad ,
\omega_{\alpha\beta}=\tilde{B}_{[\alpha\beta]}
\end{equation}
are expansion rate, shear tensor and vorticity respectively. In
general relativity the expansion rate $\theta$ can also be given by
the relation
\begin{equation}
\theta=k_{~~;\alpha}^{\alpha}
\end{equation}
and it is not dependent on the auxiliary null field $N^{\alpha}$.
The Raychaudhuri equation for the congruences of null geodesics
which gives the evolution of the volume expansion rate $\theta$ is
\begin{equation}
\frac{d\theta}{d\lambda}=-\frac{1}{2}\theta^{2}-\sigma^{\alpha\beta}\sigma_{\alpha\beta}+\omega^{\alpha\beta}\omega_{\alpha\beta}
-R_{\alpha\beta}k^{\alpha}k^{\beta}
\end{equation}
where $R_{\alpha\beta}$ is the Ricci tensor. similar equations could
be found for shear and vorticity tensors
\begin{equation}
\frac{d\sigma_{\alpha\beta}}{d\lambda}=-\theta\,\sigma_{\alpha\beta}-\tilde{C}_{\alpha\beta\gamma\rho}k^{\gamma}k^{\rho}
\end{equation}
and
\begin{equation}
\frac{d\omega_{\alpha\beta}}{d\lambda}=-\theta\,\omega_{\alpha\beta}
\end{equation}
where $C_{\alpha\beta\gamma\rho}$ is the Weyl tensor.

The extension of the equation (9) to spacetimes with non-vanishing
torsion has been previously studied in several works
\cite{Wanas,Kar}. More recently, a thorough analysis of this problem
was undertaken in \cite{Luz}. Here we briefly describe the main
result of that paper. The interested reader could find detailed
derivation of the equations there. The important notion is that in
the presence of torsion, the connecting tensor field
$B_{\alpha\beta}$ should now be defined by the following equation
instead of equation (4)
\begin{equation}
B^{~\alpha}_{\beta}=k^{\alpha}_{~~;\beta}+2\,T_{\gamma\beta}^{~~\alpha}\,k^{\gamma}
\end{equation}

In the presence of torsion, the most general form of the
Raychaudhuri equation for null geodesics is given by \cite{Luz}

$$\frac{D\theta}{d\lambda}=-R_{\alpha\beta}k^{\alpha}k^{\beta}-\left(\frac{1}{2}\theta^{2}+\sigma_{\alpha\beta}\sigma^{\alpha\beta}-\omega_{\alpha\beta}\omega^{\beta\alpha}\right)$$
$$+2T_{\rho\alpha}{}^{\beta}k^{\rho}\left(\frac{h_{\beta}{}^{\alpha}}{2}\theta+\sigma_{\beta}{}^{\alpha}+\omega_{\beta}{}^{\alpha}\right)+2k^{\gamma}\nabla_{\gamma}\left(T_{\rho}k^{\rho}\right)$$
$$+2\nabla_{\alpha}\left(T^{\alpha}\,_{\gamma\rho}k^{\gamma}k^{\rho}\right)+4\left(h^{\beta\gamma}T_{\alpha\beta\gamma}k^{\alpha}-T_{\alpha}k^{\alpha}\right)^{2}$$
$$+4T_{\mu}{}^{\alpha\beta}k^{\mu}\left[T_{\delta\beta\alpha}k^{\delta}-h_{\beta}^{\rho}T_{\delta\rho\alpha}k^{\delta}\right]$$
$$+T^{\alpha}\,_{\mu\nu}k^{\mu}\left(\widetilde{h}^{\nu\beta}-g^{\nu\beta}\right)\left(4B_{\parallel\alpha\beta}+6B_{\parallel\beta\alpha}\right)$$
\begin{equation}
-2B_{\parallel\delta}{}^{\alpha}B_{\parallel\beta\alpha}\left(\widetilde{h}^{\delta\beta}-g^{\delta\beta}\right)\,,
\end{equation}
where
$$\tilde{B}_{\alpha\beta} =\nabla_{\alpha}k_{\beta}-2T_{\alpha\gamma\beta}k^{\gamma}-2T_{\alpha\gamma\sigma}k^{\gamma}k^{\sigma}N_{\beta}+2T_{\gamma\sigma\beta}k_{\alpha}k^{\gamma}N^{\sigma}$$
$$+2T_{\gamma\rho\sigma}k_{\alpha}k^{\gamma}k^{\sigma}N_{\beta}N^{\rho}+k_{\alpha}N^{\gamma}\nabla_{\gamma}k_{\beta}+k_{\beta}N^{\gamma}\nabla_{\alpha}k_{\gamma}$$
$$-2T_{\alpha\gamma\sigma}k_{\beta}k^{\gamma}N^{\sigma}+k_{\alpha}k_{\beta}N^{\gamma}N^{\sigma}\nabla_{\sigma}k_{\gamma}$$
$$+2T_{\beta\gamma\sigma}k^{\gamma}k^{\sigma}N_{\alpha}+2T_{\gamma\sigma\rho}k_{\alpha}k_{\beta}k^{\gamma}N^{\rho}N^{\sigma}$$
\begin{equation}
+2k_{\beta}\xi_{\alpha}\xi^{\sigma}T_{\sigma\gamma\rho}k^{\gamma}k^{\rho}
\end{equation}

$$B_{\parallel\alpha\beta} \equiv B_{\alpha\beta}-\tilde{B}_{\alpha\beta}$$$$=
-2T_{\rho\sigma\beta}k^{\rho}k_{\alpha}N^{\sigma}-k_{\beta}N^{\rho}\nabla_{\alpha}k_{\rho}+2T_{\alpha\rho\sigma}k^{\rho}k_{\beta}N^{\sigma}$$
$$-2T_{\rho\sigma\gamma}k^{\rho}k_{\alpha}k_{\beta}N^{\sigma}N^{\gamma}+2T_{\alpha\rho\sigma}k^{\rho}k^{\sigma}N_{\beta}$$
$$-2T_{\rho\gamma\sigma}k^{\rho}k^{\sigma}k_{\alpha}N^{\gamma}N_{\beta}-2N^{\sigma}\xi_{\alpha}k_{\beta}T_{\sigma\gamma\rho}k^{\gamma}k^{\rho}$$
$$-2N_{\alpha}S_{\beta\gamma\sigma}k^{\gamma}k^{\sigma}-k_{\alpha}N^{\rho}\nabla_{\rho}k_{\beta}$$
\begin{equation}
-k_{\alpha}k_{\beta}N^{\rho}N^{\sigma}\nabla_{\sigma}k_{\rho}
\end{equation}
In the next section we use these equations to study light
propagation in the Poincar{\'e} gauge theory of gravity. It has been
shown in \cite{Puetz,Yasskin,Nom} particles without intrinsic
hypermomentum follow, as in general relativity, geodesics of the
metric connection. So even in the presence of torsion, light rays
follow null geodesics.
\section{Brief review of Poincar{\'e} gauge theory of gravity}
In PGT the gravitational field is described by both curvature and
torsion tensors. These in turn can be expressed in terms of tetrad
and spin connection as
 \begin{equation}
  T_{\mu\nu}{}^{i} =2(\partial_{[\mu}e_{\nu]}^{~~i})
  +\Gamma_{[\mu|j }^{i} e_{|\nu]}^{j}),
 \end{equation}

 \begin{equation}
  R_{\mu\nu i}^{~~~~j} =2(\partial_{[\mu}\Gamma_{\nu]i}^{~~~j}
  +\Gamma_{[\mu|k}^{~~~~j} \Gamma_{|\nu]i}^{~~~~k}),
\end{equation}
where $e_{~\mu}^{i}$ is the tetrad field and
\begin{equation}
g_{\mu\nu}=\eta_{ij}e_{~\mu}^{i}e_{~\nu}^{j},
\end{equation}
is the spacetime metric. The spin connection is related to the usual
holonomic connection by the relation
\begin{equation}
\Gamma^{~~j}_{\mu
i}=e_{~i}^{\nu}e^{~j}_{\rho}\Gamma^{~~~\rho}_{\mu\nu}+e_{~i}^{\nu}\partial_{\mu}e^{~j}_{\nu}
\end{equation}
We can also define the contorsion tensor as the difference between
the general connection of PGT and the Levi-Civita connection of
general relativity
\begin{equation}
K^{~~~\rho}_{\mu\nu}=\Gamma^{~~~\rho}_{\mu\nu}-\Gamma^{~~~\rho}_{\mu\nu
(0)}
\end{equation}
Where the $(0)$ subscript denotes the Christoffel symbols of general
relativity. Here the Greek indices refer to holonomic coordinate
bases and the Latin indices refer to the Local Lorentz frame. The
most general Lagrangian of the theory is a quadratic function built
by irreducible decompositions of curvature and torsion. Here we
assume a Lagrangian in the form
\begin{equation}
L_{g}=c_{1}T_{ijk}T^{ijk}+c_{2}T_{ijk}T^{jik}+c_{3}T_{i}T^{i}+c_{4}R+c_{5}R^{2}
\end{equation}
The field equations then is given by the variation of the Lagrangian
with respect to the tetrad and spin connection fields and have the
general form \cite{Shie}
\begin{eqnarray}
  \nabla_{\nu}H_{i}^{\mu\nu}-E_{i}^{~\mu}&=&{\cal T}_{i}^{~\mu},\\
  \nabla_{\nu}H_{ij}^{~~\mu\nu}-E_{ij}
  ^{~~\mu}&=&S_{ij}^{~~\mu},
 \end{eqnarray}
 where
 \begin{eqnarray}
  H_{i}^{~~\mu\nu}&:=&{\partial e L_{\rm G}\over \partial\partial_{\nu} e_{\mu}^i}
  =2{\partial e L_{\rm G}\over \partial T_{\nu\mu}{}^i},\\
  H_{ij}{}^{\mu\nu}&:=&{\partial e L_{\rm G}\over
   \partial\partial_{\nu}\Gamma_{\mu}^{~ij}}
  =2{\partial e L_g\over \partial R_{\nu\mu}{}^{ij}},
 \end{eqnarray}
 and
 \begin{eqnarray}
  E_{i} {}^{\mu}&:=&e^{\mu}{}_{i} e L_{\rm G}-T_{i \nu}{}^{j} H_{j} {}^{\nu\mu}
  -R_{i\nu}{}^{jk}H_{jk}{}^{\nu\mu},\\
  E_{ij}{}^{\mu}&:=&H_{[ij]}{}^{\mu}.
 \end{eqnarray}
The source terms here are energy-momentum and spin density tensors
respectively and are defined by
\begin{eqnarray}
 {\cal T}_{i}{}^{\mu}&:=&\frac{\partial eL_{\rm M}}{\partial e_{\mu}{}^i},\\
S_{ij}{}^{\mu}&:=&
    \frac{\partial eL_{\rm M}}{\partial \Gamma_{\mu}{}^{ij}},
\end{eqnarray}
where $L_{M}$ is the matter Lagrangian and $e$ is the determinant of
the tetrad. In a Friedman-Robertson-Walker geometry, the dual basis
or tetrad takes the following form
\begin{equation}
\vartheta^{0}=dt \quad ,\quad
\vartheta^{A}=a(t)(1+\frac{1}{4}kr^2)^{-1}dx^{A}
\end{equation}
$$
(k=0,\pm1) \quad ,\quad (A=1,2,3)$$ which gives the usual FRW metric
\begin{eqnarray}
 {\rm d}s^2 = -{\rm d}t^2+a^2(t)\Bigl[\frac{{\rm d}r^2}{1-kr^2}
 +r^2({\rm d}\theta^2+\sin^2\theta{\rm d}\phi^2)\Bigr]
\end{eqnarray}

the assumption of homogeneity and isotropy leads to following form
for the torsion and spin density tensors
\cite{Tsam,BAUERLE,Goen,Kudin}
\begin{equation}
h(t)=T_{110}=T_{220}=T_{330}=-T_{i0i}
\end{equation}
\begin{equation}
f(t)=T_{123}=T_{312}=T_{231}=-T_{[123]}
\end{equation}
and
\begin{equation}
q(t)=S_{011}=S_{022}=S_{033}=-S_{i0i}
\end{equation}
\begin{equation}
s(t)=S_{123}=S_{312}=S_{231}=-S_{[123]}
\end{equation}
with the rest of the components being zero. Due to cosmological
principle,  $q$, $s$, $h$ and $f$ can only depend on time. By
assuming that the energy-momentum tensor of the FRW universe has the
form of a perfect fluid with energy density $\rho$ and pressure $p$
and substituting the above relations in the field equations (22-23),
we get the explicit form of field equations for seven unknown
functions $q$, $s$, $h$ , $f$ , $\rho$ , $p$ and the scale factor
$a(t)$. These equations together with the equation of state for the
perfect fluid $p=\omega\rho$ and an exact model for describing the
spin density tensor, will give us the complete closed system of
equations in this case. For the  model of the spin density tensor,
one assumes a universe filled with unpolarized particles of spin
$\frac{1}{2}$, then using the averaging procedure given in
references \cite{Gasp,Nurgaliev} we have
\begin{equation}
q^2+s^2= \frac{1}{48}\hbar^{2}
A^{-2/(1+\omega)}_{\omega}\rho^{2/(1+\omega)}
\end{equation}
where $\hbar$ is the reduced Planck constant, $\omega$ is the
equation of state of the perfect fluid and $A_{\omega}$ is a
dimensional constant depending on $\omega$.

\section{Light propagation in Poincar{\'e} gauge theory of gravity}
We employ the well known techniques of geometric optics
approximation \cite{MTW}. In this limit the solution describing an
monochromatic electromagnetic wave can be described by the vector
potential $A^{\mu}$ in the Lorentz gauge as follows
\begin{equation}
A^{\mu}= \Re [A\,\epsilon^{\mu}e^{iS}] \quad \quad , \quad
A^{\mu}_{~~;\mu}=0
\end{equation}
where $\epsilon^{\mu}$ is the polarization vector and the
propagation vector $k_{\mu}$ is related to $S$ by the relation
$k_{\mu}=\partial_{\mu}S$. They also satisfy the following relations
\begin{equation}
\epsilon_{\mu}\epsilon^{\mu}=1 \quad \quad , \quad
k^{\mu}\epsilon_{\mu}=0
\end{equation}

Let us now see how the expansion, shear and vorticity change in the
presence of torsion. Using equations (7) and (12) we have
\begin{eqnarray}
\theta&=& \nabla_{\mu}k^{\mu}+2T_{~~~~\alpha}^{\beta\alpha}k^{\beta}\\
\sigma_{\alpha\beta}&=&\sigma_{\alpha\beta (0)}+2
h^{\gamma}_{\alpha}h^{\rho}_{\beta}K^{~~~~\mu}_{(\gamma\rho)}k_{\mu}\\
\omega_{\alpha\beta}&=&\omega_{\alpha\beta (0)}+2
h^{\gamma}_{\alpha}h^{\rho}_{\beta}K^{~~~~\mu}_{[\gamma\rho]}k_{\mu}
\end{eqnarray}
Where as before the $(0)$ subscripts denotes general relativistic
quantities. As one can see from eq. (39), the expansion scalar is
independent of the explicit form of the connection of the spacetime.
This is expected since the expansion scalar can also be defined with
the help of the Lie derivative and hence, is independent of the form
of the connection. This also can be seen from computing $\theta$ by
contracting eq. (14). One also can find the relation between
$\theta$ and the 'eikonal' function $S$ in the presence of torsion
by the help of relation (39). The quantities $\theta$ and
$\sigma=\frac{1}{2}\sigma_{\alpha\beta}\sigma^{\alpha\beta}$ are
called optical scalars. They play an important role in relativistic
geometric optics and are also immensely important in the study of
gravitational lensing phenomena. If we choose the affine parameter
along the null geodesics to be the conformal time $\lambda=\tau$,
Raychaudhuri equation for the FRW cosmological background with
tetrad and torsion in the forms of (32-35) then takes the form

$$\dot{\theta}
=\frac{1}{a}\Big[-\frac{1}{2}\theta_{(0)}^{2}-\sigma_{(0)}^{\alpha\beta}\sigma_{\alpha\beta
(0)}+\omega_{(0)}^{\alpha\beta}\omega_{\alpha\beta (0)}
-R_{\alpha\beta
(0)}k^{\alpha}k^{\beta}\Big]$$
\begin{equation}
-3\frac{\dot{h}}{a^2}+6\frac{\dot{a}h}{a^{3}}
\end{equation}
where a dot denotes differentiation with respect to the cosmic time
$t$. Let us analyze the effects of the last two terms in the R.H.S
of the above equation. The effects of spin and torsion where
considerable in the early universe where the density of matter where
very high. However, one should expect that these effects will be
negligible at late times when the spin density approaches zero. As a
result the $\dot{h}$ in the above equation would most probably be
negative in sign and the overall effect of the $-3\dot{h}$ term
would be to 'diverge' the beams. Also on an expanding universe,
provided that the torsion function $h$ is positive, the term
$6\frac{\dot{a}h}{a^{3}}=\frac{H}{a^{2}}$ is positive and the
overall effect of the last term would be to 'diverge' the beams
further. In summary both of the two terms involving the effects of
torsion cause divergence of neighboring light rays. Because of the
diverging effect of these last two terms, this also means that,
assuming that the null energy condition holds
$R_{\alpha\beta}k^{\alpha}k^{\beta}>0$ , one still can not easily
prove the focusing theorem which is valid in general relativity.
Equation (43) also have important consequences for measuring
cosmological distances. The expansion scalar $\theta$ is related to
the rate of change of the cross-sectional area $A$ of an
infinitesimal beam of light rays
\begin{equation}
2\theta=\frac{dA}{Ad\tau}
\end{equation}
The angular diameter distance $d_{A}$ is then related to the
expansion scalar by the relation \cite{Sasaki,SEF}
\begin{equation}
\frac{d}{d\tau}\ln(d_{A})=\frac{1}{2}\theta
\end{equation}
So one of the consequences of the modified Raychaudhuri equation is
that estimation of the angular diameter distance for distant
cosmological sources will also change in the presence of torsion.
\begin{figure*}
\centering
\begin{minipage}[b]{.4\textwidth}
{\includegraphics[scale=0.35]{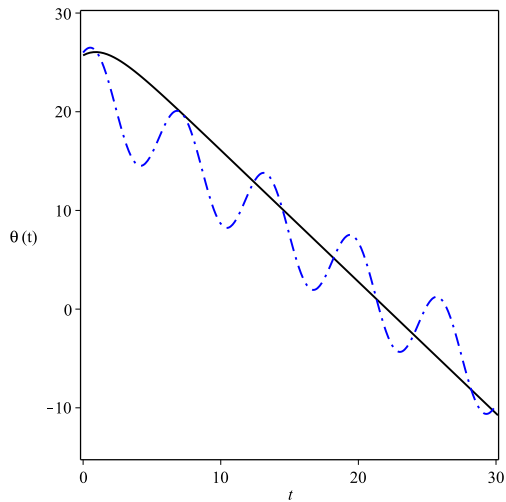}}
\end{minipage}
\caption{\small{Qualitative behavior of the expansion scalar
$\theta(t)$ as function of time for a typical choice of parameters
for a beam of light which is initially diverging ($\theta(0)>0$).
The solid black line shows the evolution for case 1 and blue dotted
line for case 2. In both cases the effects of torsion causes
$\theta(t)$ to vanish at some finite time and will be negative after
that. If $\theta$ reaches $-\infty$ at some point along the beam
trajectories, the beam starts converging after that time. }}
\end{figure*}

Remarkably the FRW torsion in the form of (32-33) does not change
the shear tensor at all as the extra terms in the second term of
R.H.S of (41) cancel out each other. By using equation (10) the
evolution equation for the shear scalar $\sigma$ takes the form
\begin{equation}
\dot{\sigma}_{(0)}+2\theta_{(0)}\,\sigma_{(0)}=0
\end{equation}
The vorticity changes as
\begin{equation}
\omega_{\alpha\beta}=\omega_{\alpha\beta (0)}-\frac{f(t)}{a(t)}
\end{equation}

We are interested in the evolution of optical scalars. The equations
(43) and (46) can be solved simultaneously to determine $\theta$ and
$\sigma$. In reference \cite{DLP} a modified form of the Frobenius'
theorem in the presence of torsion has been derived and it has been
shown that in presence of torsion, hypersurface orthogonality of
$k^{\mu}$ does not imply a vanishing vorticity tensor
$\omega_{\alpha\beta}$ . However, in the special case of the
homogenous and isotropic torsion (32-33) used here, the extra terms
in eq. (B5) of reference \cite{DLP} are canceled and one can set the
vorticity to zero in this case. In order to solve the system of
equations, the explicit form of the scale factor $a(t)$ and torsion
function $h(t)$ are needed which one can get by solving the PGT
field equations. The most general case involving seven unknown
functions $q$, $s$, $h$ , $f$ , $\rho$ , $p$ and $a(t)$ is very
complicated but there are several very interesting special cases
which we will present here. For a full discussion of the solutions
of the PGT field equations in various special cases see references
\cite{Goen,AQK,Mink1,Mink2,Ao}.

\subsection{Case 1}
For an illustrative example we consider the solutions to the PGT
field equations with zero spin density and axial torsion but
non-vanishing vector torsion. In this case we set $f=s=q=0$ which
represent reflection invariant solutions. The solution to the field
equation also depends on the specific choice of coefficients
$c_{1}-c_{5}$ in the Lagrangian (21) and also the value of $k$ in
the metric (31). If $c_{1}+3c_{3}\neq 0$ then we have a solution
with an effective cosmological constant in the form of
$\Lambda=9c_{5}/(c_{1}+3c_{3})$. For the case of $\Lambda >0$ and
$k=0$, the solutions are as follows
\begin{equation}
a(t)=a_{0}e^{\sqrt{\frac{\Lambda}{3}}t}
\end{equation}
\begin{equation}
h(t)=\small{-\frac{\sqrt{3\Lambda}}{9}\frac{a(c\sqrt{3\Lambda}+3)e^{-c\sqrt{\frac{\Lambda}{3}}t}+b(d\sqrt{3\Lambda}
+3)e^{-d\sqrt{\frac{\Lambda}{3}}t}}{ae^{-c\sqrt{\frac{\Lambda}{3}}t}+be^{-d\sqrt{\frac{\Lambda}{3}}t}}}
\end{equation}
where $a$ and $b$ are constants of integration and $c$ and $d$ are
some combination of coefficients. The solution for the scale factor
in this case show a de Sitter expansion for the universe. Replacing
this solutions to evolution equations for the optical scalars (42)
and (45) and solving the system of equations, we find

$$\theta(t)=3a_{0}\sqrt{\frac{\Lambda}{3}}+\Bigg[(c+d+3)Ei\Bigg(1,\frac{(c+d+3)e^{-2\sqrt{\frac{\Lambda}{3}t}}}{2a^{2}_{0}}\Bigg)$$
$$+2\Lambda(c+d+3)^{3}e^{-2\sqrt{\frac{\Lambda}{3}t}}-2\Lambda(c+d+3)^{2}t$$$$+\Big(27+4\Lambda
(c+d+3)^{2}a^{2}_{0}\Big)e^{-\frac{1}{2}\frac{(c+d+3)^3}{a^{2}_{0}}e^{-2\sqrt{\frac{\Lambda}{3}t}}}\Bigg]$$
\begin{equation}
\times \Bigg[\sqrt{\Lambda}(2c+2d+6)\Bigg]^{-1}
\end{equation}
Where $Ei(a,z) = z^{(a-1)}\Gamma(1-a,z)$ is the exponential
integral. In the above relation, the first term on the R.H.S is the
general relativistic result while all the other terms represent
corrections due to the effects of torsion.
$$\sigma(t)=-\frac{21\Lambda}{16}(c+d+3)e^{-2\sqrt{\frac{\Lambda}{3}t}}+\frac{3}{70}Ei\Big(1,
\frac{e^{-2\sqrt{\frac{\Lambda}{3}t}}}{4}\Big)$$
$$-\frac{(c+d+3)^{2}}{50}t^{2}+\frac{\sqrt{3\Lambda}}{32}\Bigg(t\,Ei\Big(1, \frac{1}{4}t\Big)-4e^{-\sqrt{\frac{3\Lambda}{4}}t}\Bigg)$$
The evolution of the expansion scalar $\theta$ in this case is shown
in figure 1 for a typical choice of parameters (solid black line).
We assumed that $\theta(0)>0$ which means that the beam is initially
diverging. As one can see, $\theta$ will reach zero at some finite
point in the affine parameter and will be negative after that. This
means that the initially diverging beam will start converging at
some point if $\theta$ reaches $-\infty$ at some finite affine
parameter point along the trajectory of null geodesics. One can also
find a relation for the angular diameter distance in this case by
using relations (44) and (49).

\subsection{Case 2}
For the specific case of $c_{1}=2$ and $c_{5}> 0$ there is another
possible interesting solution of PGT field equations as follows\\

$$a(t)=$$
\begin{equation}
\small{a_{0}\exp\Big[\frac{-\mu\,h_{0}\omega\,t+3H_{0}\omega\,t+\mu\,h_{0}\sin(\omega\,t)+3\alpha\,R_{0}\cos(\omega\,t)}{3\omega}\Big]}
\end{equation}
and
\begin{equation}
h(t)=-\beta^{-1}R_{0}\sin(\omega\,t)+h_{0}\cos(\omega\,t)
\end{equation}
where $\mu=c_{1}-c_{4}$, $\omega=\sqrt{\frac{2\mu}{c_{1}c_{5}}}$,
 $\alpha=\sqrt{\frac{c_{5}\mu}{72c_{1}}}$, $\beta=\sqrt{\frac{8\mu
c_{1}}{c_{5}}}$ and $h_{0}$ and $H_{0}$ and $R_{0}$ are initial
values of the torsion scalar $h$, the Hubble parameter and the
curvature scalar respectively. In this case the optical scalars
$\theta$ and $\sigma$ are given by solving the system of equations
(42) and (45) as
\begin{eqnarray}
\theta(t) =
C_{1}e^{-t}\nonumber\\+\small{\Bigg[\Big(-18H_{0}\beta\,\omega^{2}-6\beta
R_{0}\alpha\omega
-2\beta h_{0}\mu+18R_{0}\omega\Big)\cos(\omega\,t)}\nonumber\\
+\Big(18H_{0}\beta\,\omega-6\beta R_{0}\alpha
-2\beta h_{0}\mu+18R_{0}\omega^{2}\Big)\sin(\omega\,t)\nonumber\\
\small{+6\Big(\frac{1}{3}\mu
h_{0}-H_{0}-\frac{1}{2}\Big)(1+\omega^{2})\beta\Bigg]\times
\Bigg[6\beta(\omega^{2}+1)\Bigg]^{-1}}
\end{eqnarray}
the shear scalar in this case then is given by
\begin{equation}
\sigma(t)=C_{2}e^{\int \frac{\theta(t)a(t)}{C_{3}}dt}
\end{equation}

where $C_{1-3}$ again are constants of integration. The behavior of
the expansion scalar in this case is also depicted in figure 2 by
the blue dotted line. Similar to case 1, the expansion scalar for an
initially diverging beam reaches zero at some point along the ray
trajectories and will be negative after that.

Let us now briefly discuss how the presence of torsion modify the
evolution of the polarization vector defined in (37). It is a well
known fact that in general relativity, the polarization vector is
parallel transported along the trajectory of light rays. However it
has been shown that in a spacetime with torsion this will not be the
case in general \cite{Ritis}. The Maxwell's wave equation in the
presence of torsion reads \cite{Ritis,Pras}

$$\nabla_{\nu}\nabla^{\nu}A^{\mu}-(2T^{\mu\nu\rho}+Q^{\mu\nu\rho})(\nabla_{\nu}A_{\rho}-\nabla_{\nu}A_{\rho})$$
\begin{equation}
-(2\nabla_{\nu}T_{~~\rho}^{\mu\nu}+2\nabla_{\nu}Q_{~~\rho}^{\mu\nu}+R_{~~\rho}^{\mu})A\rho=0
\end{equation}
where
\begin{equation}
Q^{~~\rho}_{\mu\nu}\equiv
T^{~~\rho}_{\mu\nu}+\delta^{\rho}_{\mu}T^{~~\sigma}_{\nu\sigma}-\delta^{\rho}_{\nu}T^{~~\sigma}_{\mu\sigma}
\end{equation}
Using equation (37) we get the following equation for the evolution
of the polarization vector

$$k^{\nu}\nabla_{\nu}\epsilon^{\mu}=\frac{1}{2}\epsilon^{\mu}\Big[(2T^{\rho\nu\sigma}+Q^{\sigma\nu\rho})(\epsilon_{\rho}^{*}k_{[\sigma}\epsilon_{\nu]}+
\epsilon_{\rho}k_{[\sigma}\epsilon^{~~*}_{\nu]})$$
\begin{equation}
-\frac{1}{2}(2T^{\rho\nu\sigma}+Q^{\sigma\nu\rho})(k_{\sigma}\epsilon_{\nu}+
k_{\sigma}\epsilon_{\nu})\Big]
\end{equation}
substituting for torsion in the FRW background from equation (32)
and (33) we see that the polarization vector is not parallel
transported  along the light ray trajectories.
\section{Conclusion and discussion}
Geometric optics and the evolution of the optical scalars have found
many interesting cosmological applications over the years. Any
slight deviation from the general relativistic equations governing
the light propagation, may have far reaching consequences in many
areas of astrophysics and cosmology. For example, number and
location of images from cosmological gravitational lenses,
development of caustics, measurement of the angular diameter
distance for cosmological objects and also study of the CMB spectrum
are all directly dependent on the light propagation equations and
the properties of the bundle of light rays. In Poincar{\'e} gauge
theory of gravity, the gravitational interaction is described not
only by curvature, but also by the torsion of spacetime. Torsion is
also present in many other approaches to gravity such as string
theory \cite{GSW}, supergravity \cite{PT} and theories involving
twistors \cite{Pap}. It has been shown that the presence of torsion
will modify the Raychaudhuri equation for the expansion scalar and
also equations governing the evolution of shear and vorticity
tensors. Here, we derive the explicit form of the expansion and
shear for a homogeneous and isotropic FRW background in the
Poincar{\'e} gauge theory of gravity. We also solved the evolution
equations for optical scalars and found the analytical solutions for
two different cases in PGT. The result shows that in general, the
evolution of optical scalars in a cosmological background will be
different in spacetimes with torsion and in general relativity. This
may provide an indirect test of torsion theories by carefully
studying gravitational lensing and other similar effects.

\end{document}